\documentclass[twocolumn,floats,floatfix,aps,pra]{revtex4}
\usepackage{amsfonts,amssymb,amsmath}
\usepackage{color,calc}
\usepackage[dvips]{graphicx}
\usepackage{bm}

\def\be{ \begin{equation} }
\def\ee{ \end{equation} }
\def\bea{ \begin{eqnarray} }
\def\eea{ \end{eqnarray} }
\def\bse{ \begin{subequations} }
\def\ese{ \end{subequations} }
\def\ba{ \begin{array} }
\def\ea{ \end{array} }
\def\bt{\begin{tabular}}
\def\et{\end{tabular}}

\def\U{\mathbf{U}}

\def\H{\mathbf{H}}

\newcommand{\ket}[1]{\vert #1\rangle}

\def\P{\Omega_p}

\def\p{p}
\def\q{q}

\def\P{P}
\def\P{\mathcal{P}}

\def\prob{\mathcal{P}}

\DeclareMathAlphabet\mathbfcal{OMS}{cmsy}{b}{n}

\begin{document}


\title{Quantum sensing of weak electric and magnetic fields by coherent amplification of energy level shift effects}

\author{Nikolay V. Vitanov}

\affiliation{Department of Physics, St. Kliment Ohridski University of Sofia, 5 James Bourchier blvd., 1164 Sofia, Bulgaria}

\date{\today }

\begin{abstract}
A method for measuring small energy level shifts in a qubit by coherent amplification of their effect is proposed.
It is based on the repeated application of the same interaction pulse in two manners: with the same phase of each subsequent pulse, and with an alternating phase shift of $\pi$ (i.e. a minus sign) from pulse to pulse.
Two specific types of pulses are considered: a resonant $\pi$ pulse and an adiabatic chirped pulse, both of which produce complete population inversion with high fidelity.
In the presence of a weak ambient external electric or magnetic field, the ensuing Stark or Zeeman shift leads to an energy level shift and hence a static detuning.
In both the resonant and adiabatic approaches, a small level shift does not alter the transition probability very much; however, it can significantly change the dynamical phases in the propagator.
The repeated application of the same pulse
 greatly amplifies the changes in the dynamical phases and maps them onto the populations.
Hence the effect of the level shift can be measured with good accuracy.
It is found that sequences of pulses with alternating phases deliver much greater error amplification and much steeper excitation profiles around resonance, thereby providing much higher sensitivity to small energy level shifts.
Explicit analytic estimates of the sensitivity are derived using the well-known non-crossing Rosen-Zener and Rabi models and the level-crossing Demkov-Kunike model.
This recipe provides a simple tool for rapid and accurate sensing of weak electric and magnetic fields by using the same pulse generating an inversion quantum gate, without sophisticated tomography or entangling operations.
\end{abstract}

\maketitle


\section{Introduction}\label{Sec:intro}
%


In scalable quantum computation \cite{Nielsen2000}, the quantum gates in a quantum circuit have to be implemented with very high fidelity, with the admissible error in the range of $10^{-3}$ to $10^{-4}$, depending on the quantum error correction protocol \cite{Lidar2013}.
The current state of the art in trapped-ions experiments features errors of $10^{-5}$ \cite{Brown2011,Gaebler2016} and even $10^{-6}$ \cite{Harty2014} for single-qubit gates, and errors of the order of $10^{-3}$ \cite{Ballance2016,Gaebler2016} for two-qubit gates.
The detrimental cross-talk to neighboring ion qubits has been suppressed to values of $10^{-5}$ \cite{Piltz2014} and $10^{-6}$ \cite{Craik2017}, and the errors in ion transport have been reduced below $10^{-5}$ too \cite{Kaufmann2018}.
In superconducting qubits, fidelities of 99.9\% for single-qubit and 99.4\% for two-qubit gates have been reported \cite{Barends2014}.
Various steps in qubit readout and quantum gate tomography, e.g. population shelving \cite{Stevens1998,Sorensen2006,Moller2007,Schafer2020}, have to be implemented with very high fidelity too.


Such tiny errors require very good control of the driving field as well as the environment, e.g. compensation of ambient electric and magnetic fields to a very high degree.
This is crucial in a quantum circuit where the entire propagator of the particular gate must be very well controlled, i.e. both the probability and the propagator phases have to be very stable.
The phases, in particular, are very sensitive to a detuning shift, which can emerge due to both fluctuating frequency of the laser, microwave or radiofrequency generator, and energy level shifts caused by uncompensated magnetic or electric fields.

In order to characterize such unwanted detuning shifts in the course of the measurement it is very useful to have a simple, fast and reliable method to detect and measure such shifts.
Measuring the populations after the application of a single pulse, which generates the quantum gate, is not reliable because the populations are fairly insensitive to detuning shifts, whereas the gate phases, which are prone to these shifts, are invisible to such a measurement.
Performing full quantum tomography would reveal the changes in the gate phases but such a tomography could be very time-consuming.

To this end, I propose in this paper a conceptually very simple technique for sensing and measuring small detuning shifts.
It is based on the repeated application of the same gate-generating pulse, which greatly amplifies the effect of the detuning shift due to quantum interference and maps it onto the populations.
In this manner, one does not need any change in the experimental apparatus (i.e. the pulse amplitude, duration, shape, frequency and possibly chirp), and avoids the introduction of additional experimental parameters beyond the gate-generating pulse.
In this manner, the proposed technique is simpler than other, more sophisticated quantum sensing methods \cite{Degen2016} and quantum gate tomography \cite{Merkel2013}.
I consider the implementation of this sensing method by applying the pulses used in two major quantum control techniques for population inversion: a resonant $\pi$ pulse and an adiabatic chirped pulse.

The paper is organized as follows.
The concept of the sensing method is introduced in Sec.~\ref{Sec:concept} and the general features of the transition probability generated the two types of sensing sequences are presented in Sec.~\ref{Sec:multi-pass}.
Three exactly soluble analytic models are presented in Secs.~\ref{Sec:RZ}, \ref{Sec:Rabi}, and \ref{Sec:DK}.
The last section \ref{Sec:conclusions} wraps up the results and presents some discussion on the limitations of the method and an outlook of its possible extensions.

%


\section{Concept \label{Sec:concept}}

\begin{figure}[t]
\begin{tabular}{cc}
\bt{c} \includegraphics[width=0.22\columnwidth]{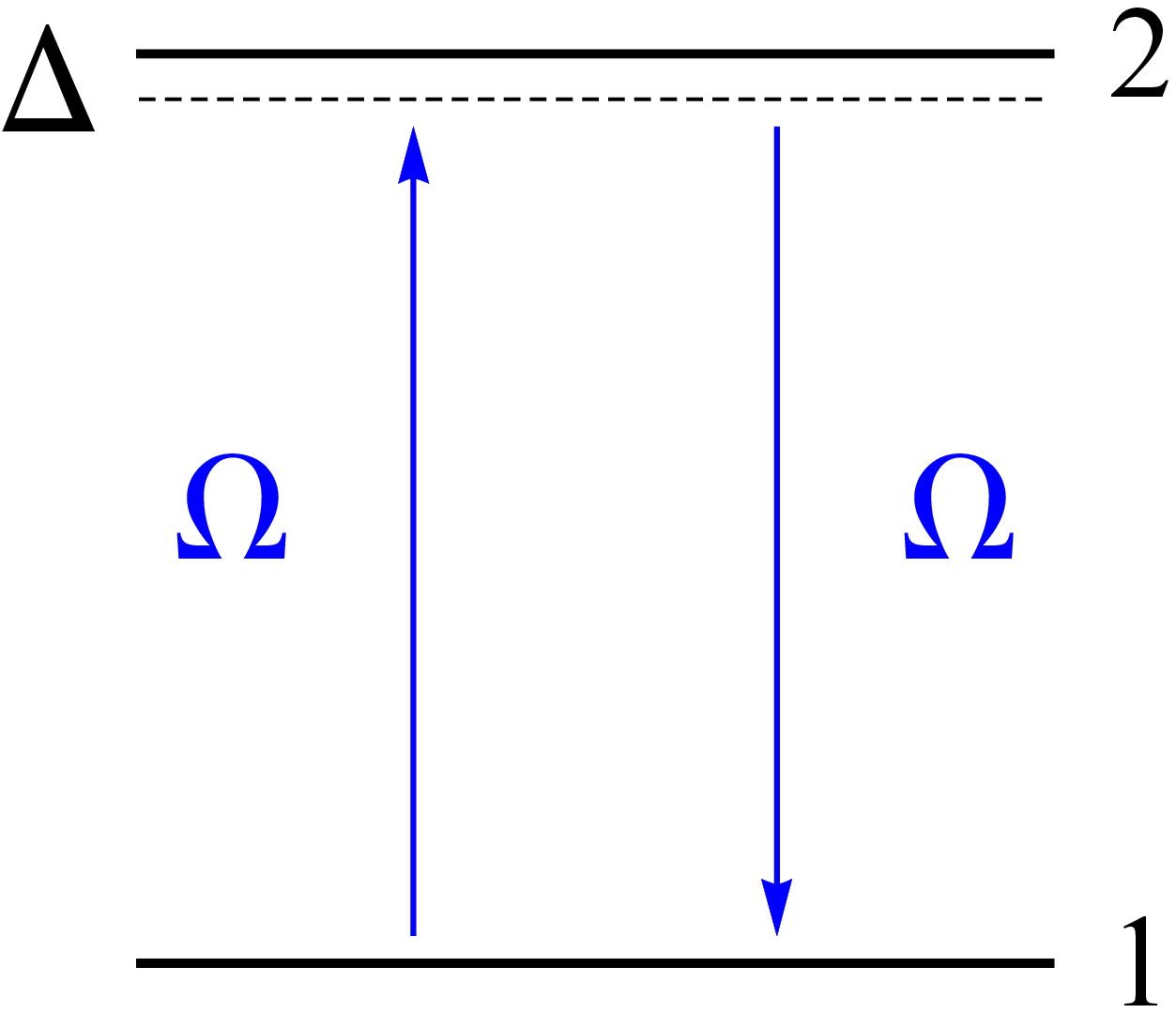} \et & \bt{c} \includegraphics[width=0.75\columnwidth]{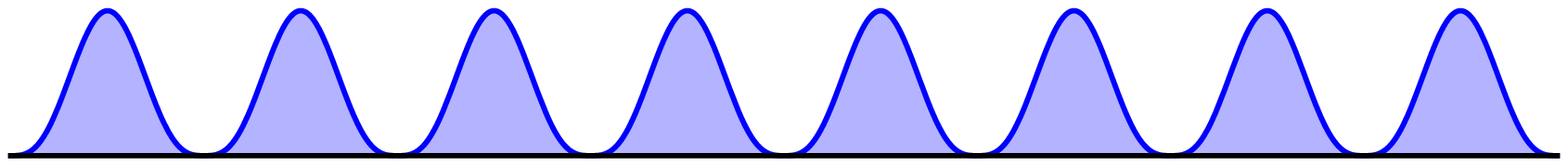} \et \\ \\
\bt{c} \includegraphics[width=0.22\columnwidth]{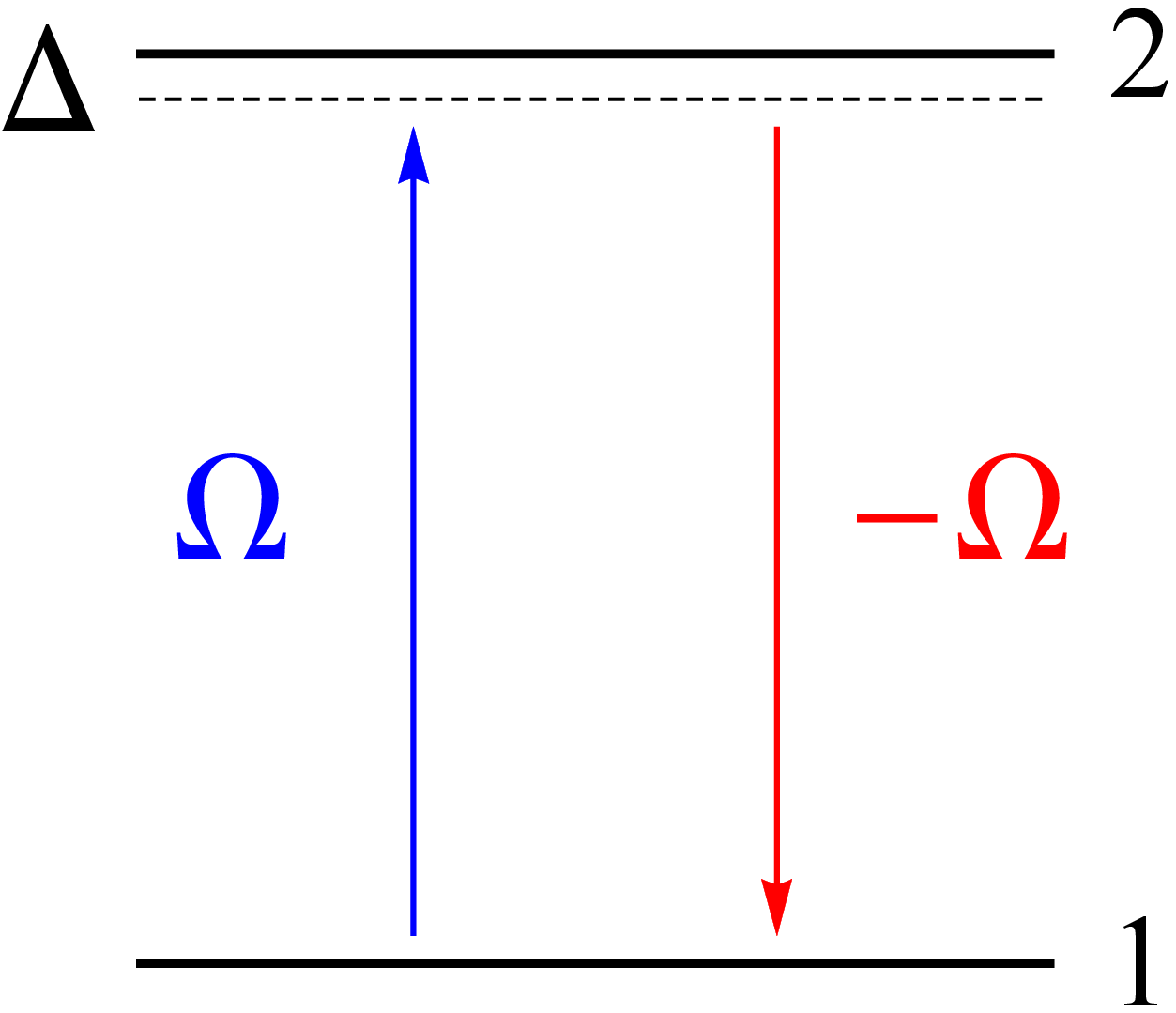} \et & \bt{c} \includegraphics[width=0.75\columnwidth]{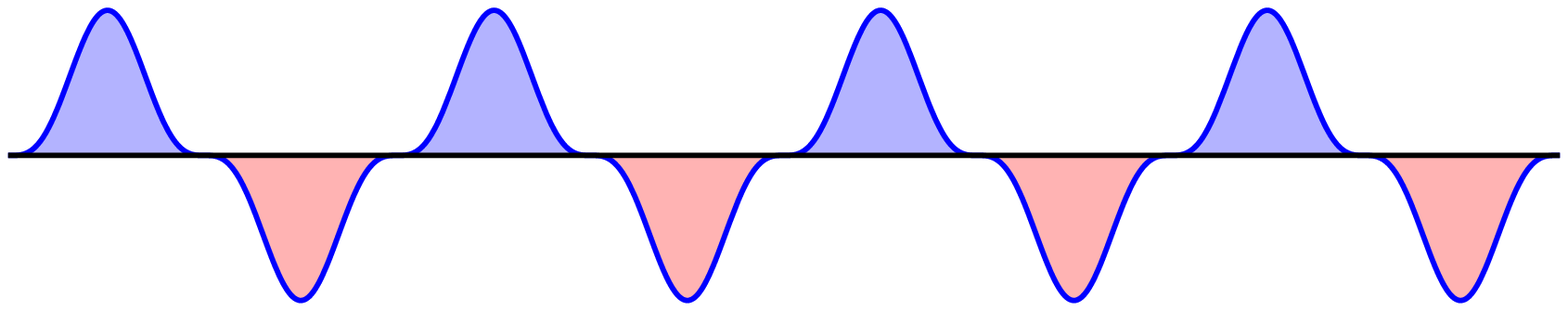} \et \\
& \bt{c} \includegraphics[width=0.52\columnwidth]{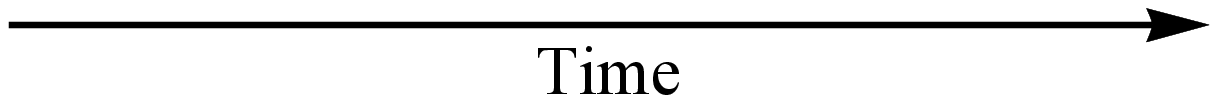} \et
\end{tabular}
\caption{
Pulse sequences used to measure a small detuning shift $\Delta_0$ from resonance by repeating the single-pulse interaction, while keeping its phase the same (top), or flipping it by $\pi$ from pulse to pulse (bottom).
}
\label{fig:2ss}
\end{figure}

I consider two types of pulse sequences applied to a two-state quantum system, see Fig.~\ref{fig:2ss}.
In one of them, the same pulse is applied $N$ times, see Fig.~\ref{fig:2ss} (top), and then the populations are measured.
In the other, the same pulse is again applied $N$ times but the phase of every other pulse is shifted by $\pi$, i.e. a minus sign is applied to the Rabi frequency of every even-numbered pulse in the sequence, see Fig.~\ref{fig:2ss} (bottom).
%
The sequences may contain an odd or even number of pulses.
On exact resonance it is assumed that each pulse produces complete (for resonant pulse) or almost complete (for adiabatic chirped pulse) population inversion.
This is achieved by a $\pi$ pulse in the former case and a chirped pulse with a pulse area of a few $\pi$ in the latter.

Because the excitation profile has its maximum value (and hence a vanishing first-order derivative versus the detuning) at resonance, a small detuning shift has a very little effect on the populations.
However, such a detuning shift changes the dynamic phases of the propagator much more significantly.
If one is restricted to measuring populations, as it is assumed here, these phase shifts are invisible in a single-pulse interaction; however, they are mapped onto the populations by the interference generated by a train of pulses.
These pulses are chosen to be identical to the single $\pi$ pulse, or the single chirped adiabatic pulse, mentioned above, with the only difference being the sign flip of the Rabi frequency in the sign-alternating sequence.
In this manner, no new parameters are introduced in addition to the dynamic parameters of the single-pulse propagator (the transition probability and the two dynamic phases).
This fact is important because, in the absence of other uncertainties, the population changes are directly linked to the detuning shift.

The rationale for the two pulse sequences considered here is the following.
It might appear at first sight that repeating the same pulse, without phase shifts, is the most natural approach to coherent error amplification.
Indeed, this case is carefully analyzed here.
However, it turns out that a sequence of pulses with alternating Rabi frequency signs is a far better approach as far as detuning shift sensing is concerned.
When considered more carefully, this is readily understood: a pair of two resonant pulses with a $\pi$ phase shift cancel each other's effect exactly because the propagator for the second pulse is the Hermitean conjugate of the propagator of the first, thereby resulting in the identity operation.
In other words, a pair of resonant pulses with Rabi frequencies $\Omega$ and $-\Omega$ will produce no overall change and exactly zero transition probability.
A small detuning will break this symmetry and the cancellation will not occur, giving rise to a nonzero transition probability.
When such a pair is repeated $N$ times, the nonzero probability is coherently amplified very quickly.

Furthermore, sequences of both even and odd number of pulses $N$ are used.
With the transition probability for zero static detuning being equal to one, or almost one, one should keep in mind that the multi-pass transition probability for an odd $N$ is equal to one, or almost one, while it is zero, or almost zero, for an even $N$.
Hence the sensing feature around the zero static detuning appears as a \emph{spike} for odd $N$ and a \emph{dip} for even $N$.
It is the \emph{width} of this feature and its \emph{slope}, which are important for the frequency shift sensing rather than whether it is a spike or a dip.

In order to not only sense an energy level shift but accurately measure it using the pulse sequence scenario, an analytic relation between the single-pulse and $N$-pulse probabilities is needed.
Such a relation for a SU(2) propagator is available \cite{Vitanov1995} and it has been used recently for quantum gate tomography \cite{Vitanov2018,Vitanov2020}.
Here it is used extensively, in combination with explicit analytic formulas for the propagator for three popular exactly soluble two-state models, the non-crossing Rosen-Zener \cite{Rosen1932,Vitanov1994} and Rabi \cite{Shore1990} models and the level-crossing Demkov-Kunike model \cite{Demkov1969,Hioe1985,Zakrzewski1985,Suominen1992}.
They allow one to explicitly link the $N$-pulse transition probability to the detuning shift.

\section{Multi-pass transition probability}\label{Sec:multi-pass}
%

The Hamiltonian of a coherently driven two-state quantum system, in the rotating-wave approximation \cite{Shore1990}, reads
\be\label{H-2}
\H(t) = \tfrac12 \left[\begin{array}{cc} -\Delta(t) & \Omega(t) \\ \Omega(t) & \Delta(t) \end{array}\right],
\ee
where $\Delta(t)$ is the system-field frequency detuning and $\Omega(t)$ is the Rabi frequency of the coupling between the two states.
The Rabi frequency is supposed to be pulse-shaped and the detuning may contain a static part $\Delta_0$ and a chirp,
\be
\Delta (t) = \Delta_0 + \beta f(t).
\ee
In this paper, the objective is to measure a static error in the detuning around zero, i.e. the static detuning $\Delta_0$ will be the (unknown) quantity to be determined.

For arbitrary $\Omega(t)$ and $\Delta(t)$, the propagator corresponding to the traceless Hamiltonian \eqref{H-2} is a SU(2) matrix, which is expressed in terms of the complex-valued Cayley-Klein parameters $a$ and $b$ ($|a|^2 +|b|^2=1$) as
\be\label{U}
\U = \left[\begin{array}{cc} a & -b^* \\ b & a^* \end{array}\right].
\ee
If the system is initially in state $\ket{1}$, the probabilities for remaining in state $\ket{1}$ and for transition to state $\ket{2}$ are
\be\label{2ss-pq}
\q = |a|^2,\quad \p = |b|^2. 
\ee
with $p+q = 1$.
%
In the following sections, the connections between $\Omega(t)$ and $\Delta(t)$ and the Cayley-Klein parameters will be explicitly presented for three specific analytically soluble models.

Instead of the two complex Cayley-Klein parameters $a$ and $b$, the propagator \eqref{U} can be expressed in terms of three real parameters: the transition probability $p$ and the dynamical phases $\xi$ and $\eta$ (sometimes called St\"uckelberg phases),
\be\label{propagator}
\U = \left[\begin{array}{cc} e^{i \xi} \sqrt{1-p} & - e^{-i\eta} \sqrt{p}  \\ e^{i\eta} \sqrt{p} & e^{-i \xi} \sqrt{1-p} \end{array}\right] ,
\ee
where $a = e^{i \xi} \sqrt{1-p}$ and $b = e^{i\eta} \sqrt{p}$.

\subsection{Sequence of pulses with the same phases\label{Sec:plus}}

In order to determine the populations after $N$ passes we need to find the $N$-pass propagator $\mathbfcal{U}_N = \U^{N}$.
It has been proved \cite{Vitanov1995} that the $N$-th power of any SU(2) propagator, parameterized as in Eq.~\eqref{U}, reads
\be\label{U-N}
\mathbfcal{U}_N = \left[\begin{array}{cc}
\cos N\theta + ia_i \dfrac{\sin N\theta}{\sin \theta} & -b^* \dfrac{\sin N\theta}{\sin \theta} \\
 b \dfrac{\sin N\theta}{\sin \theta} &  \cos N\theta  - ia_i \dfrac{\sin N\theta}{\sin \theta}
\end{array}\right],
\ee
where $a = a_r + i a_i$ and
\be\label{theta}
\theta = \arccos (a_r) \quad (0 \leqq \theta \leqq \pi).
\ee
Therefore, the transition probability after $N$ passes is
\be\label{P-N}
\prob_N =  p\, \frac{\sin^2 (N\theta)}{\sin^2 (\theta)}.
\ee
For $N=2$, after simple algebra one finds
\be
\U_{\Omega,\Delta} \U_{\Omega,\Delta} = \left[\begin{array}{cc} a^2 - |b|^2 & -2 b^* a_r \\ 2 b a_r & (a^*)^2 - |b|^2 \end{array}\right]. \label{2ss-UppU}
\ee
The double-pass transition probability is equal to $\prob_2 = 4 p (1-p) \cos^2 \xi$.

\subsection{Sequence of pulses with alternating phases\label{Sec:minus}}

For the pulse sequence with alternating phases we need a modification of the above result as follows.
Consider a second interaction with the same magnitudes of $\Omega(t)$ and $\Delta(t)$, but with the opposite sign of $\Omega(t)$, see Fig.~\ref{fig:2ss}.
The respective propagator can be obtained from Eq.~\eqref{U} by simple algebraic operations \cite{Vitanov1999nonlinear,Vitanov2018,Vitanov2020} and, very importantly, can be expressed with the \emph{same} Cayley-Klein parameters $a$ and $b$,
\be\label{U-}
\U_{-\Omega,\Delta} = \left[\begin{array}{cc} a & b^* \\ -b & a^* \end{array}\right].
\ee
The respective double-pass propagator reads
\be
\U_{-\Omega,\Delta} \U_{\Omega,\Delta} = \left[\begin{array}{cc} a^2 + |b|^2 & -2 i b^* a_i \\ -2 i b a_i & (a^*)^2 + |b|^2 \end{array}\right], \label{2ss-UmpU} 
\ee
where $\U_{\Omega,\Delta}$ is the same as $\U$ of Eq.~\eqref{U}.
The double-pass transition probability is equal to $\prob_2^\pm = 4 p (1-p) \sin^2 \xi$, which is different from the double-pass transition probability of $\prob_2 = 4 p (1-p) \cos^2 \xi$ in the case of the same phases due to the St\"uckelberg phase $\xi$.
This difference will be greatly amplified in the $N$-pass probabilities.

For the sign-alternating sequence, we can use Eq.~\eqref{2ss-UmpU} as the basic building block and derive the $2n$-pass propagator using the connection between Eqs.~\eqref{U} and \eqref{U-N} above.
Obviously, the parameter $a$ is now replaced by $a^2 + |b|^2 = 
  1 -2 a_i^2 + 2i a_r a_i  $, the parameter $b$ is replaced by $-2 i b a_i$, and the parameter $\theta$ is redefined as
\be\label{Theta}
\Theta = 
\arccos (1-2a_i^2).
\ee
Then the propagator for the sign-alternating sequence of $2N$ pulses reads
\be\label{U2-N}
\mathbfcal{U}_{2n}^\pm = \left[\begin{array}{cc}
\cos n\Theta + 2i a_r a_i \dfrac{\sin n\Theta}{\sin \Theta} & -2 i b^* a_i \dfrac{\sin n\Theta}{\sin \Theta} \\
 -2 i b a_i \dfrac{\sin n\Theta}{\sin \Theta} &  \cos n\Theta  - 2i a_r a_i \dfrac{\sin n\Theta}{\sin \Theta}
\end{array}\right].
\ee
The propagator for $2n+1$ pulses can be obtained by multiplying $\mathbfcal{U}_{2n}^\pm$ by $\U$ of Eq.~\eqref{U},
\be\label{U2-Nodd}
\mathbfcal{U}_{2n+1}^\pm = \U \mathbfcal{U}_{2n}^\pm = \left[\begin{array}{cc}
a C + 2i a_i S & - b^* C \\
b C & a^* C - 2i a_i S
\end{array}\right],
\ee
with
\be
C= \dfrac{\cos (n+\frac12)\Theta}{\cos \frac12\Theta},\quad S = \dfrac{\sin n\Theta}{\sin \Theta}.
\ee
Therefore, the transition probability after $2n$ and $2n+1$ pulses reads
\bse\label{P-N-pm}
\begin{align}
\P_{2n} &= 
 p\, \frac{\sin^2 n\Theta}{\cos^2 \frac12\Theta}, \label{Ppme}
\\
\P_{2n+1} &= p\, \frac{\cos^2 (n+\frac12)\Theta}{\cos^2 \frac12\Theta} .  \label{Ppmo}
\end{align}
\ese
In the derivation of Eq.~\eqref{Ppme}, it is used that $a_i^2 = \sin^2 \frac12\Theta$, which follows from Eq.~\eqref{Theta}.

In the limit of nearly complete population transfer by a single pulse, which is concerned here, we have $|b| \approx 1$ and $|a| \ll 1$.
Then $|a_r| \ll 1$ and $|a_i| \ll 1$ and hence $\theta \approx \pi/2$ and $\Theta \ll 1$.
More accurately, we have
\bse
\begin{align}
\theta &\approx \tfrac12\pi - a_r - \tfrac16 a_r^3 + \cdots, \\
\Theta &\approx 2|a_i| + \tfrac13 |a_i|^3 + \cdots
\end{align}
\ese
Therefore, for sequences of pulses of the same phase,
\bse\label{PN++}
\begin{align}
\P_{2n} &\approx (2n)^2 a_r^2, \label{PN++even} \\
\P_{2n+1} &\approx 1 - a_i^2 - (2n+1)^2 a_r^2, \label{PN++odd}
\end{align}
\ese
while for sequences of pulses of alternating phases,
\bse\label{PN+-}
\begin{align}
\P_{2n}^\pm &\approx (2n)^2 a_i^2, \label{PN+-even} \\
\P_{2n+1}^\pm &\approx 1 - (2n+1)^2 a_i^2, \label{PN+-odd}
\end{align}
\ese
where the relation $p = 1 - a_r^2 -a_i^2$ has been accounted for.
While retaining the dominant asymptotic terms, especially in Eq.~\eqref{PN++odd}, it has been also used that $a_i$ is of order $O(\delta)$ and $a_r$ is of order $O(\delta^2)$.
Obviously, the difference between the two types of sequences is that
for a sequence of pulses with the same phases, the $N$-pulse transition probability \eqref{P-N} is controlled by the real part $a_r$ of the diagonal Cayley-Klein parameter $a$,
while for a sequence of pulses with alternating phases, the $N$-pulse transition probability is controlled by the imaginary part $a_i$ of $a$.
As we will see, this is an important difference and  these observations show the significance of the properties of the parameter $a$.

\subsection{Case studies \label{Sec:models}}

In order to examine the performance of the general formulas \eqref{PN++} and \eqref{PN+-} in regard to sensing, we consider three
analytically soluble models, which provide explicit analytic expressions for the parameters $a_r$ and $a_i$, from which one can find out the dependence of the populations on the static detuning shift $\Delta_0$.
The first model is the Rosen-Zener model, which assumes a bell-shaped hyperbolic-secant pulse shape and a static detuning.
It is suitable for studying the sensitivity of a population inverting resonant $\pi$ pulse to a small detuning shift.
The second model is the Rabi model, which assumes a rectangular pulse shape and a constant detuning.
It is of the same type as, and simpler than the Rosen-Zener model.
The comparison of the two models reveals the effects of the sharp pulse edges in the Rabi model.
The third model is the Demkov-Kunike model, which assumes the same hyperbolic-secant pulse shape as the Rosen-Zener model but the detuning is a sum of a hyperbolic-tangent-shaped detuning and a static detuning.
It is suitable for modelling adiabatic passage via a level crossing in the presence of a small detuning shift.

\begin{figure}[t]
\includegraphics[width=0.96\columnwidth]{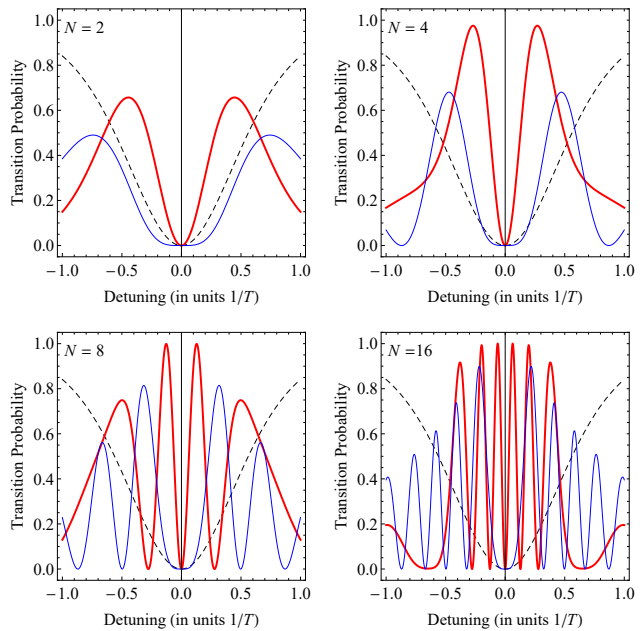}
\caption{
Transition probability vs the static detuning shift $\Delta_0$ for sequences of $N=2$, 4, 8, and 16 identical pulses indicated by the boxed numbers.
The pulse shape is hyperbolic secant, with a pulse area of $\pi$ and width $T$, and the detuning is constant (Rosen-Zener model).
In each frame, the transition probability is plotted for a sequence of pulses with the same phase (thin blue solid curve), and a sequence of pulses with alternating phases (thick red solid curve).
The dashed curve shows the no-transition probability for a single pulse and serves as a reference.
}
\label{fig:rz-even}
\end{figure}

\section{Resonant $\pi$ pulses: Rosen-Zener model}\label{Sec:RZ}

The non-crossing \emph{Rosen-Zener model} is defined as \cite{Rosen1932}
\be\label{RZ model}
\Omega(t) = \Omega_0\, \text{sech}\, (t/T),\quad \Delta(t) = \Delta_0.
\ee
The Cayley-Klein parameters of the Rosen-Zener propagator are  \cite{Rosen1932,Vitanov1994,Vitanov2012}
\bse
\begin{align}
a &= \frac{\Gamma(\nu) \Gamma(\nu-\lambda-\mu)} { \Gamma(\nu-\lambda) \Gamma(\nu-\mu)} , \\
b &= -i \frac{\sin(\pi \alpha / 2)}{\cosh(\pi \delta / 2)},
\end{align}
\ese
where $\Gamma(z)$ is Euler's gamma function \cite{wolfram} and
\be
\lambda = \alpha/2, \quad \mu = -\alpha/2, \quad \nu = (1+i\delta)/2,
\ee
with $\alpha = \Omega_0 T$ and $\delta = \Delta_0 T$.
For $\Delta_0 = 0$, the Rosen-Zener model describes a resonant pulse of pulse area $\pi \Omega_0 T = \pi \alpha$.
Therefore, a resonant $\pi$ pulse is realized with $\alpha = 1$.
The presence of the static detuning $\Delta_0$ allows to simulate the effect of a detuning shift on resonant excitation.

Of particular interest in the present context is the behavior of the parameter $a$ for small detuning.
For a $\pi$ pulse ($\alpha=1$) we have
\be\label{a-rz}
a \approx i \frac{\pi\delta}{2} + \pi\delta^2\ln2 - i \pi \delta^3 \left[ \frac{\pi^2}{24} + (\ln2)^2 \right] + O(\delta^4).
\ee
From here, it follows that
\bse
\begin{align}
\theta &\approx \frac \pi 2 - \pi\delta^2\ln2 + O(\delta^4), \\
\Theta &\approx \pi \delta + O(\delta^3).
\end{align}
\ese

\begin{figure}[t]
\includegraphics[width=0.96\columnwidth]{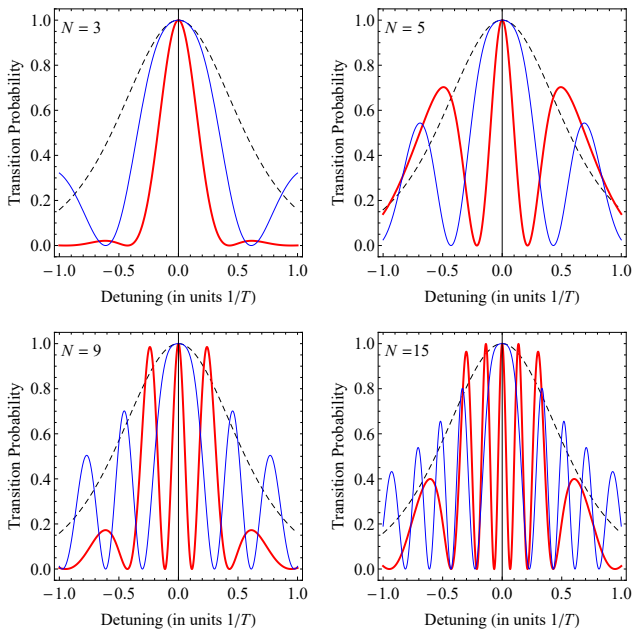}
\caption{
The same as Fig.~\ref{fig:rz-even} but for an odd number of pulses, $N=3$, 5, 9, and 15.
Here the dashed curve shows the transition probability for a single pulse and serves as a reference.
}
\label{fig:rz-odd}
\end{figure}

Figure \ref{fig:rz-even} shows the transition probability vs the static detuning  shift $\Delta_0$ for sequences of an even number of hyperbolic-secant pulses, each with a pulse area of $\pi$, and Fig.~\ref{fig:rz-odd} shows the transition probability for sequences of an odd number of pulses.
The probabilities are calculated from Eqs.~\eqref{P-N} and \eqref{P-N-pm}.
Around resonance, a narrow feature forms which gets more narrow as the number of pulses $N$ increases.
For an even number of pulses, the feature shows up as a dip (Fig.~\ref{fig:rz-even}), and for an odd number of pulses, it appears as a spike (Fig.~\ref{fig:rz-odd}).
This feature is much more narrow for sequences of alternating phases (thick curves) than for sequences of the same phases (thin curves).

Using Eqs.~\eqref{PN++}, \eqref{PN+-} and \eqref{a-rz}, it is easy to derive the behavior of the transition probability for $|\delta|\ll 1$.
For a sequence of pulses with the same phases, the approximation is
\bse
\begin{align}\label{Peven-p-a}
P_{2n} &\approx (2n \pi \ln2)^2  \delta^4 , 
\\
\label{Podd-p-a}
P_{2n+1} &\approx 1 - \tfrac14 \pi^2 \delta^2 +  \left[\tfrac{1}{24}\pi^2 - 4n(n+1) (\ln2)^2 \right] \pi^2\delta ^4 . 
\end{align}
\ese
For a sequence of pulses of alternating phases, it reads
\bse\label{P-m-a}
\begin{align}\label{Peven-m-a}
P_{2n} & \approx n^2 \pi^2 \delta^2 , 
\\
\label{Podd-m-a}
P_{2n+1} & \approx 1  - (n+\tfrac12)^2 \pi^2 \delta^2. 
\end{align}
\ese

\begin{figure}[t]
\includegraphics[width=0.75\columnwidth]{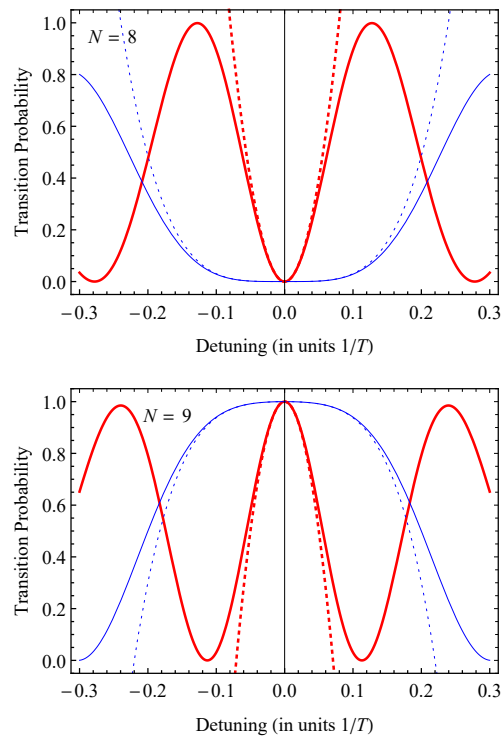}
\caption{
Transition probability vs the static detuning shift $\Delta_0$ for sequences of $N=8$ and 9 identical pulses for sech pulses with a pulse area of $\pi$ and width $T$
 (Rosen-Zener model).
The two frames are zoomed-in and modified versions of the corresponding frames in Figs.~\ref{fig:rz-even} and \ref{fig:rz-odd}.
In each frame, the transition probability is plotted for a sequence of pulses with the same phase (thin blue solid curve), and a sequence of pulses with alternating phases (thick red solid curve).
The dashed curves show the approximations for $|\delta|\ll 1$: Eqs.~\eqref{Peven-p-a} and \eqref{Peven-m-a} for $N=8$ pulses, and Eqs.~\eqref{Podd-p-a} and \eqref{Podd-m-a} for $N=9$ pulses.
}
\label{fig:rz-approx}
\end{figure}

The four approximate formulas \eqref{Peven-p-a}-\eqref{Podd-m-a} are plotted in Fig.~\ref{fig:rz-approx} and compared to the exact values.
A very good agreement between exact and approximate values is observed for $|\delta| \ll 1$, as it should be the case.

For a sequence of pulses with alternating phases, the dependence on $N$ and $\delta$ near resonance is the same for an even and odd number of pulses, albeit inverted, see Eqs.~\eqref{Peven-m-a} and \eqref{Podd-m-a}: the transition probability departs as $\propto N^2 \delta^2$ from its resonant value.
This means that the excitation profile can be squeezed as much as desired by merely increasing the number of pulses $N$.

The behavior of the transition probability for a sequence of pulses with the same phases is rather different:  the transition probability departs as $\propto N^2 \delta^4$ from zero for an even number of pulses [Eq.~\eqref{Peven-p-a}], while for an odd number of pulses, its departure from unity is described by a sum of a $N$-independent term $\propto \delta^2$ and a term $\propto N^2 \delta^4$ [Eq.~\eqref{Podd-p-a}].
Consequently, the transition probability is much flatter in the range near resonance than for sequences of pulses with alternating phases.
Moreover, the presence of the $N$-independent term $\propto \delta^2$ for odd $N$ impedes the squeezing of the excitation profile by increasing $N$.
In either cases, due to the different departure law from resonance, the profile is much broader than for alternating-phase sequences, as indeed seen in Figs.~\ref{fig:rz-even}-\ref{fig:rz-approx}.

The conclusion is that, as far as squeezing of the excitation profile near resonance is concerned, the sequences of pulses with alternating phases outperform the sequences of pulses with equal phases and therefore, are much more efficient for sensing of small detuning shifts.
Moreover, the simple and accurate approximate formulas, Eqs.~\eqref{Peven-p-a}-\eqref{Podd-m-a}, allow ones to not only sense a detuning shift but also measure it by measuring populations.
Any ambiguity, which may arise for a larger detuning shift due to the multiple oscillations for large $N$, can be resolved be making measurements for different $N$.

In order to estimate the sensitivity of this technique, consider the sequences with alternating phases, for which the deviation from the resonant value is, in the lowest order, $(N\pi\delta/2)^2$, see Eqs.~\eqref{P-m-a}.
By setting this deviation to $\frac12$ and recalling that $\delta = \Delta_0 T$, we find that the half-width-at-half-maximum of the spike or the dip is
\be\label{delta-rz}
\Delta_{\frac12} = \frac{\sqrt{2}}{N\pi T} \approx \frac{0.45}{N T} .
\ee
Therefore, the sensitivity can be increased by increasing $T$ (which is the well-known Fourier bandwidth argument) or by increasing $N$.
For example, a sequence of ten $\pi$ pulses with alternating phases, each of duration 10 $\mu$s, allows one to sense and measure a detuning shift of 4.5 kHz.
If, instead of a deviation of $\frac12$, we can measure a population deviation of $\frac1{10}$, then the sensitivity of the same arrangement improves to 2 kHz.

As it is clear from the approximations \eqref{Peven-p-a} and \eqref{Podd-p-a}, the excitation profile can be squeezed by sequences of pulses with the same phases too.
However, the scaling is much less efficient, $\Delta_{\frac12} \propto 1/N^{\frac12}$, rather than $1/N$.
By setting the deviation to $\frac12$, we find from Eq.~\eqref{Peven-p-a} that
\be
\Delta_{\frac12} = \frac{1}{\sqrt{N\pi \sqrt{2}\, \ln2}\, T} \approx  \frac{0.57}{\sqrt{N}\, T}.
\ee
Hence a sequence of ten $\pi$ pulses with the same phases, each of duration 10 $\mu$s, allows one to sense and measure a detuning shift of 18 kHz, a factor of 4 larger than for sequences of alternating phases.
Therefore, by merely flipping the phase of every other pulse in the sequence, one can achieve much stronger (quadratically enhanced) squeezing and hence much better sensitivity.

\section{Rectangular pulses: Rabi model}\label{Sec:Rabi}

One can apply the same approach using pulses of rectangular shape, to the so-called Rabi model,
\be\label{Rabi model}
\Omega(t) = \left\{ \begin{array}{cc} \Omega_0, & |t|\leqq T/2 \\ 0, & |t|> T/2 \end{array} \right. ,
\quad \Delta(t) = \Delta_0.
\ee
The Cayley-Klein parameters of the Rabi propagator are far simpler than for the Rosen-Zener model \cite{Shore1990},
\bse
\begin{align}
a &= \cos \left(\tfrac{1}{2} \sqrt{\alpha ^2+\delta ^2}\right)+\frac{i \delta  \sin \left(\frac{1}{2} \sqrt{\alpha ^2+\delta ^2}\right)}{\sqrt{\alpha ^2+\delta ^2}} , \\
b &= -\frac{i \alpha  \sin \left(\frac{1}{2} \sqrt{\alpha ^2+\delta ^2}\right)}{\sqrt{\alpha ^2+\delta ^2}},
\end{align}
\ese
with $\alpha = \Omega_0 T$ and $\delta = \Delta_0 T$.
Note that the pulse area is equal to $\alpha$; therefore, a resonant $\pi$ pulse is realized with $\alpha = \pi$.
For such a $\pi$ pulse, the Taylor expansion of the Cayley-Klein parameter $a$ reads
\be
a \approx i\frac{\delta}{\pi} - \frac{\delta^2}{4\pi} - i \frac{\delta^3}{2\pi^3} + \frac{\delta^4}{16\pi^3}  + O(\delta^5).
\ee
Hence to the lowest order in $\delta$ we have $a_r = -\delta^2/(4\pi)$ and $a_i = \delta/\pi$.
The transition probability for the pulse sequences with alternating phases can be calculated from Eqs.~\eqref{P-N-pm}.
For small $\delta$ we find from Eqs.~\eqref{PN+-}
\bse\label{PN+-Rabi}
\begin{align}
\P_{2n}^\pm &\approx (2n)^2 \frac{\delta^2}{\pi^2}, \label{PN+-even-Rabi} \\
\P_{2n+1}^\pm &\approx 1 - (2n+1)^2 \frac{\delta^2}{\pi^2}, \label{PN+-odd-Rabi}
\end{align}
\ese
By setting the transition probability to $\frac12$ and recalling that $\delta = \Delta_0 T$, we find for both odd and even $N$
\be\label{delta-rabi}
\Delta_{\frac12} \approx \frac{\pi}{\sqrt{2}\, N T} \approx \frac{2.22}{N T} .
\ee
This value is a factor of about 5 larger than that for the Rosen-Zener model, Eq.~\eqref{delta-rz}.

\begin{figure}[t]
\includegraphics[width=0.70\columnwidth]{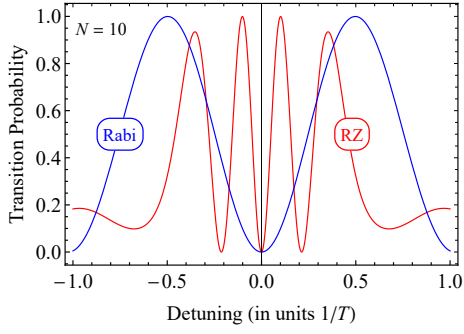}
\caption{
Comparison of the Rosen-Zener and Rabi models for sequences of 10 pulses with alternating phases.
The sech and rectangular pulse have the same peak amplitude and the same area $\pi$.
}
\label{fig:rabi-rz}
\end{figure}

Figure \ref{fig:rabi-rz} compares the excitation profiles for the Rabi and Rosen-Zener models generated by sequences of 10 $\pi$ pulses of alternating phases.
The pulse area of each pulse is $\pi$ and the peak Rabi frequency is the same for each model.
Obviously, the feature near resonance is much more narrow for the Rosen-Zener model (a factor of about 5, as noted above).
The physical reason is that the rectangular pulses in the Rabi model exhibit typical power broadening, due to its sharp edges, while the smooth pulse in the Rosen-Zener model has no power broadening at all \cite{Vitanov2001,Halfmann2003,Boradjiev2013}.
Note that some pulse shapes, with wings vanishing as $t^{-n}$ exhibit even \emph{power narrowing} \cite{Boradjiev2013} and they might provide even better sensitivity than sech pulses.

\section{Adiabatic chirped pulses: Demkov-Kunike model}\label{Sec:DK}

The level-crossing {Demkov-Kunike model} is defined as \cite{Demkov1969}
\be\label{DK model}
\Omega(t) = \Omega_0\, \text{sech}\, (t/T),\quad \Delta(t) = \Delta_0 + B \tanh(t/T).
\ee
For $\Delta_0=0$, the Demkov-Kunike model reduces to the Allen-Eberly-Hioe model \cite{Allen1975,Hioe1984,Silver1985}, also known as the \emph{complex-sech pulse} in NMR.
It is the most beautiful example of chirped adiabatic passage involving a level crossing.
The addition of the static detuning $\Delta_0$ allows to simulate a detuning shift in this model.
For $B=0$, the Demkov-Kunike model reduces to the Rosen-Zener model.
For $B=\Delta_0$, the Demkov-Kunike model turns into the Bambini-Berman model \cite{Bambini1981}.

The Cayley-Klein parameters for this model are \cite{Demkov1969,Vitanov2012}
\bse
\begin{align}
a &= \frac{\Gamma(\nu) \Gamma(\nu-\lambda-\mu)} { \Gamma(\nu-\lambda) \Gamma(\nu-\mu)} , \\
b &= -\frac{i\alpha \Gamma(1-\nu) \Gamma(\nu-\lambda-\mu)} {2 \Gamma(1-\lambda) \Gamma(1-\mu)},
\end{align}
\ese
where
\bse
\begin{align}
\lambda &= \left(\sqrt{\alpha^2-\beta^2} - i\beta\right) / 2, \\
\mu &= -\left(\sqrt{\alpha^2+\beta^2} + i\beta\right) / 2, \\
\nu &= (1+i\delta-i\beta) / 2,
\end{align}
\ese
with $\alpha = \Omega_0 T$, $\beta = B T$, and $\delta = \Delta_0 T$.

In the Demkov-Kunike model, we have for $\alpha=\beta=2$ (corresponding to pulse area of $2\pi$ and chirp rate of $B=2/T$)
\be\label{a-dk}
a \approx 0.086 + 0.165i\delta - 0.052\delta^2 + 0.036i\delta^3 + O(\delta^4),
\ee
from where we find
\bse
\begin{align}
\theta &\approx 1.484 + 0.052\delta ^2 + O(\delta ^4), \\
\Theta &\approx 0.33 \delta + O(\delta ^3).
\end{align}
\ese

\begin{figure}[t]
\includegraphics[width=0.96\columnwidth]{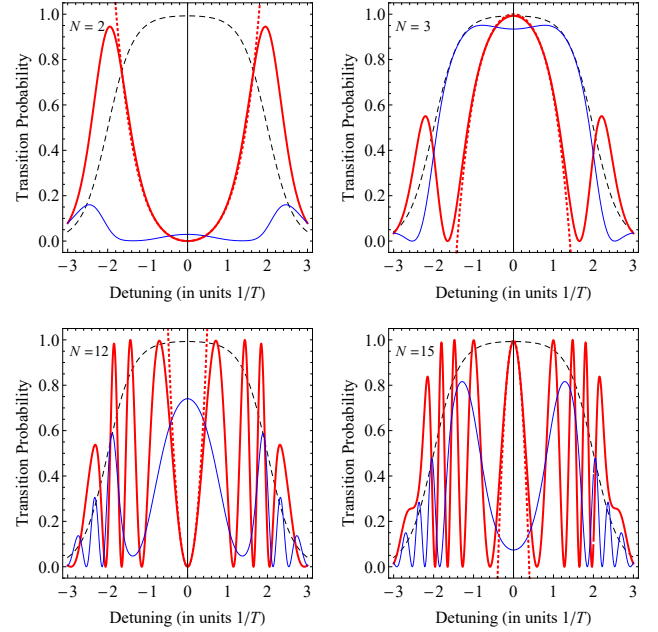}
\caption{
Transition probability in the Demkov-Kunike model vs the static detuning shift $\Delta_0$ for sequences of $N$ identical pulses indicated by the boxed numbers.
The pulse shape is hyperbolic secant, with a pulse area of $2\pi$ and width $T$, and the detuning given by a hyperbolic-tangent chirp (with $\beta = 2/T$) and a constant term $\Delta_0$.
In each frame, the transition probability is plotted for a single pulse (dashed curve), a sequence of pulses with the same phase (thin blue solid curve), and a sequence of pulses with alternating phases (thick red solid curve).
The dotted curves illustrate the approximations \eqref{Pdk-even-m-a} for even $N$ and \eqref{Pdk-odd-m-a} for odd $N$.
}
\label{fig:dk}
\end{figure}

The transition probability for the Demkov-Kunike model can be calculated from Eqs.~\eqref{P-N} and \eqref{P-N-pm} and it is plotted in Fig.~\ref{fig:dk}.
We are interested again in the feature near zero static detuning which emerges as a spike (for odd $N$) or a dip (for even $N$).
Once again, the features produced by the sequences with alternating phases are much more narrow  than the features produced by both the single pulse and the sequences of pulses with the same phases.
The single-pulse profile is much broader near zero detuning compared to the single-pulse profile of the Rosen-Zener model in Figs.~\ref{fig:rz-even} and \ref{fig:rz-odd}, which manifests the robustness characteristic of adiabatic passage methods.
Consequently, it is more difficult to squeeze the excitation profile with a small number of pulses (the frames with 2 and 3 pulses), but for longer pulse sequences the desired squeezing still occurs, especially for pulse sequences with alternating phases.
As for resonant pulses, it is possible to sense a detuning shift by using the same chirped pulse used for population inversion, without changing anything except for the sign of the Rabi frequency.

One can derive the behavior of the transition probability for small $\delta$ using Eqs.~\eqref{PN++}, \eqref{PN+-} and \eqref{a-dk}.
For a sequence of pulses with the same phases, the picture is rather messy (see Fig.~\ref{fig:dk}) and the approximation is not very meaningful.
For a sequence of pulses of alternating phases, Eqs.~\eqref{PN+-} and \eqref{a-dk} give the asymptotics $(|\delta|\ll 1)$
\bse
\begin{align}\label{Pdk-even-m-a}
P_{2n} & \approx (0.33n)^2 \delta^2 (1+0.438\delta^2) ,
\\
\label{Pdk-odd-m-a}
P_{2n+1} & \approx 1 - [0.33 (n+\tfrac12)]^2 \delta^2 (1+0.438\delta^2),
\end{align}
\ese
where higher terms in $\delta$ are retained for better accuracy.
These approximations are plotted in Fig.~\ref{fig:dk} by dashed curves.
They allow one, as in the Rosen-Zener model, to estimate the sensitivity of this technique.
By setting the transition probability to $\frac12$ and recalling that $\delta = \Delta_0 T$, we find for both odd and even $N$
\be\label{delta-dk}
\Delta_{\frac12} \approx \frac{4.3}{N T}.
\ee
For example, a sequence of ten $\pi$ pulses with alternating phases, each of duration 10 $\mu$s, allows one to sense and measure a detuning shift of 43 kHz.
This is almost a factor of 10 larger than for resonant $\pi$ pulses in the Rosen-Zener model, cf.~Eq.~\eqref{delta-rz} versus Eq.~\eqref{delta-dk}.
This is not surprising because the adiabatic passage techniques, as described here by the Demkov-Kunike model, are resilient to parameter variations, including the detuning.
For larger values of the pulse area and the chirp, the transition probability becomes even more robust to parameter errors and the sensitivity decreases even more.

\section{Conclusions\label{Sec:conclusions}}

This paper presented a method for detection and measurement of small detuning shifts generated, e.g., by weak external electric or magnetic fields.
The method uses coherent amplification of transition probability errors by a train of identical pulses in two setups: with the same phase of each subsequent pulse, and with an alternating phase shift of $\pi$ from pulse to pulse.
Two kinds of pulses were considered: a resonant $\pi$ pulse and an adiabatic chirped pulse, both of which are standard quantum control tools for complete population inversion.
In either cases, small detuning shifts do not change the transition probability very much; however, they modify the dynamical phases in the propagator much more significantly, which are amplified and mapped onto the populations by the repeated application of the same pulse.

Explicit analytic estimates were derived using the well-known non-crossing Rosen-Zener and Rabi models and the level-crossing Demkov-Kunike model.
Based on the analytical results and numerical simulations, it was concluded that sequences of pulses with alternating phases outperform those with the same phases, as far as sensing is concerned: they generate much steeper, and hence much narrower, excitation profiles around resonance, thereby providing much higher sensitivity to detuning shifts.
Smooth resonant $\pi$ pulses, exemplified by the Rosen-Zener model, are by far the best performer, with the greatest sensitivity.
Alternatively, Gaussian pulses, for which analytic results (albeit not so simple) are also available \cite{Vasilev2004}, deliver similar performance.
It is worth considering also pulses of Lorentzian-type shapes (with wings vanishing as $\propto t^{-n}$), which exhibit power narrowing \cite{Boradjiev2013} and may deliver even better sensitivity than sech pulses.
Rectangular pulses, represented by the Rabi model, exhibit broader profiles (by a factor of 5) than the Rosen-Zener model due to power broadening generated by the sharp edges of the pulse.
Chirped adiabatic pulses, exemplified by the Demkov-Kunike model considered here, are far less suitable for sensing than resonant pulses because of the inherent robustness of adiabatic techniques to parameter variations.
Similar results can be obtained for linearly chirped Gaussian pulses for which analytical results are available \cite{Vasilev2005}.
Furthermore, using rectangular pulses with a linear chirp, as in the popular Landau-Zener-St\"uckelberg-Majorana model \cite{Landau1932,Zener1932,Stueckelberg1932,Majorana1932} model in its finite version \cite{Vitanov1996}, is an inappropriate option for sensing either because it features both adiabaticity and sharp pulse edges.
Therefore, \emph{sequences of smooth resonant $\pi$ pulses with alternating phases} are identified as the most suitable for sensing of small detuning shifts.

It is worth emphasizing that the proposed sensing method uses identical pulses (except for the possible $\pi$ phase shift from pulse to pulse).
In this manner, no additional uncertainties and errors are introduced which might mask the effects of small level shifts.
The proposed technique is very convenient for practical use because the same pulse used as a NOT gate in a quantum circuit can be used to detect detuning shifts.
Hence this simple recipe provides an efficient tool for rapid sensing of weak electric and magnetic fields, without sophisticated tomography setups or entangling operations.
It is applicable to all kinds of experimental platforms wherein the environment causes energy level shifts.
Particularly promising are Rydberg atoms and ions \cite{Saffman2010,Fan2015,Mokhberi2020,Adams2020} due to their increased sensitivity to electric field variations.

The present idea is basically the opposite of the idea that the NMR community is pursuing since many years, that is, alternating pulse phases such that the detuning effects are suppressed, termed dynamical rephasing or dynamical decoupling (DD).
The simplest such sequence is the Carr-Purcell-Meiboom-Gill (CPMG) two-pulse sequence \cite{Carr1954,Meiboom1958}.
An important development is the XY-4 sequence \cite{Maudsley1986,Viola1999}, which uses 4 phase-shifted pulses.
The XY-4 sequence is used as a building block for periodic DD, in which it is applied sequentially \cite{Gullion1990}, and concatenated DD (CDD) sequences, which concatenate lower-order CDDs recursively, starting from XY-4 \cite{Khodjasteh2005, Piltz2013,Casanova2015}.
Another important development is the concept of robust DD sequences \cite{Souza2011,Genov2017}, which are resilient to various pulse errors.
Yet another development is the extension to quantum systems with more than two states \cite{Vitanov2015}.
Further details can be found in a comprehensive review \cite{Souza2012}.

It is likely that, taking inspiration from the vast DD literature, the straightforward flips in the pulse phase in the sensing method proposed here could be optimized further, by letting the relative phases from pulse to pulse be free control parameters.
However, this task is well beyond the scope of the present work.

It should be clear that the sensing method presented here is based upon quantum interference.
Therefore, the measurement should be fast enough in order to avoid noise and hence dephasing.

Finally,  in this work only detuning errors have been considered.
In a real experiment, Rabi frequency errors might occur too, which would turn the parameter estimation problem into a multi-parameter one.
Therefore, the method is strictly applicable in the absence of such errors.
However, the underlying assumption is that the $\pi$ pulses used in the sensing sequence is of very high fidelity because it is used in some quantum circuit, which would be the main experiment, i.e. the loss of probability does not come from it but from the external ambient fields.
The present method allows one to quickly measure, once in a while, whether the external field has changed.
More importantly, a closer inspection shows that small Rabi frequency errors do not affect the method as far as detuning sensing is concerned.
It is only important that the Rabi frequency does not change during the sensing sequence, even if it is slightly different from its nominal $\pi$ pulse value.

\acknowledgments

This work is supported by the 
 Bulgarian Science Fund Grant DO02/3 (Quant-ERA Project ERyQSenS).



\begin{thebibliography}{99}

\bibitem{Nielsen2000} M. A. Nielsen and I. L. Chuang, \emph{Quantum Computation and Quantum Information} (Cambridge University Press, Cambridge, U.K., 2000).


\bibitem{Lidar2013} D. Lidar and T Brun (Eds.), \emph{Quantum Error Correction} (Cambridge, Cambridge University Press, 2013).


\bibitem{Brown2011}
K. R. Brown, A. C. Wilson, Y. Colombe, C. Ospelkaus, A. M. Meier, E. Knill, D. Leibfried, and D. J. Wineland,
Phys. Rev. A \textbf{84}, 030303(R) (2011).

\bibitem{Gaebler2016}
J. P. Gaebler, T. R. Tan, Y. Lin, Y. Wan, R. Bowler, A. C. Keith, S. Glancy, K. Coakley, E. Knill, D. Leibfried, and D. J. Wineland,
Phys. Rev. Lett. \textbf{117}, 060505 (2016).

\bibitem{Harty2014}
T. P. Harty, D. T. C. Allcock, C. J. Ballance, L. Guidoni, H. A. Janacek, N. M. Linke, D. N. Stacey, and D. M. Lucas,
Phys. Rev. Lett. \textbf{113}, 220501 (2014).

\bibitem{Ballance2016}
C. J. Ballance, T. P. Harty, N. M. Linke, M. A. Sepiol, and D. M. Lucas,
Phys. Rev. Lett. \textbf{117}, 060504 (2016).

\bibitem{Piltz2014}
C. Piltz, T. Sriarunothai, A.F. Var\'on, and C. Wunderlich, Nat. Commun. \textbf{5}, 4679 (2014).

\bibitem{Craik2017}
D. P. L. Aude Craik, N. M. Linke, M. A. Sepiol, T. P. Harty, J. F. Goodwin, C. J. Ballance, D. N. Stacey, A. M. Steane, D. M. Lucas, and D. T. C. Allcock,
Phys. Rev. A \textbf{95}, 022337 (2017).

\bibitem{Kaufmann2018}
P. Kaufmann, T. F. Gloger, D. Kaufmann, M. Johanning, and C. Wunderlich,
Phys. Rev. Lett. \textbf{120}, 010501 (2018).

\bibitem{Barends2014}
R. Barends, J. Kelly, A. Megrant, A. Veitia, D. Sank, E. Jeffrey, T. C. White, J. Mutus, A. G. Fowler, B. Campbell, Y. Chen, Z. Chen, B. Chiaro, A. Dunsworth, C. Neill, P. O'Malley, P. Roushan, A. Vainsencher, J. Wenner, A. N. Korotkov, A. N. Cleland, and J. M. Martinis,
Nature \textbf{508}, 500 (2014)


\bibitem{Stevens1998}
D. Stevens, J. Brochard, and A. M. Steane, Phys. Rev. A \textbf{58}, 2750 (1998).


\bibitem{Sorensen2006}
J. L. S{\o}rensen, D. M{\o}ller, T. Iversen, J. B. Thomsen, F. Jensen, P. Staanum, D. Voigt, and M. Drewsen, New J. Phys. \textbf{8}, 261 (2006).

\bibitem{Moller2007} D. M{\o}ller, J. L. S{\o}rensen, J. B. Thomsen, and M. Drewsen, Phys. Rev. A \textbf{76}, 062321 (2007).

\bibitem{Schafer2020}
V. M. Sch\"afer, \emph{Fast Gates and Mixed-Species Entanglement with Trapped Ions}, Springer Theses (2020).

\bibitem{Degen2016} C. L. Degen, F. Reinhard, and P. Cappellaro, Rev. Mod. Phys. \textbf{89}, 035002 (2017).

\bibitem{Merkel2013}
 S. T. Merkel, J. M. Gambetta, J. A. Smolin, S. Poletto, A. D. Corcoles, B. R. Johnson, C. A. Ryan, and M. Steffen, Phys. Rev. A \textbf{87}, 062119 (2013).


\bibitem{Vitanov1995}
N. V. Vitanov and P. L. Knight, Phys. Rev. A \textbf{52}, 2245 (1995).

\bibitem{Vitanov2018} N. V. Vitanov, Phys. Rev. A \textbf{97}, 053409 (2018).

\bibitem{Vitanov2020} N. V. Vitanov, New J. Phys. \textbf{22}, 023015 (2020).

\bibitem{Rosen1932} N. Rosen and C. Zener, Phys. Rev. \textbf{40}, 502 (1932).

\bibitem{Vitanov1994} N. V. Vitanov, J. Phys. B: \emph{At. Mol. Opt. Phys.} \textbf{27}, 1351 (1994).

\bibitem{Shore1990} B. W. Shore, \emph{The Theory of Coherent Atomic Excitation} (Wiley, New York, 1990).

\bibitem{Demkov1969} Y. N. Demkov and M. Kunike, Vestn. Leningr. Univ., Ser. 4: Fiz. Khim. \textbf{16}, 39 (1969).

\bibitem{Hioe1985} F. T. Hioe and C. E. Carroll, Phys. Rev. A \textbf{32}, 1541 (1985).

\bibitem{Zakrzewski1985} J. Zakrzewski, Phys. Rev. A \textbf{32}, 3748 (1985).

\bibitem{Suominen1992} K.-A. Suominen and B. M. Garraway, Phys. Rev. A \textbf{45}, 374 (1992).

\bibitem{Vitanov1999nonlinear} N. V. Vitanov and K.-A. Suominen, Phys. Rev. A \textbf{59}, 4580 (1999).

\bibitem{Vitanov2012} N. V. Vitanov, \emph{Quantum Transitions: An Introduction to Time-Dependent Quantum Dynamics of Atoms and Molecules} (Sofia University Press, 2012).

\bibitem{wolfram} https://functions.wolfram.com

\bibitem{Vitanov2001} N. V. Vitanov, B. W. Shore, L. P. Yatsenko, K. B\"ohmer, T. Halfmann, T. Rickes, and K. Bergmann, Opt. Commun. \textbf{199}, 117 (2001).

\bibitem{Halfmann2003} T. Halfmann, T. Rickes, N. V. Vitanov, and K. Bergmann, Opt. Commun. \textbf{220}, 353 (2003).

\bibitem{Boradjiev2013} I. I. Boradjiev and N. V. Vitanov, Opt. Commun. \textbf{288}, 91 (2013).

\bibitem{Allen1975} L. Allen and J. H. Eberly, \emph{Optical Resonance and Two-Level Atoms} (Dover, New York, 1975).

\bibitem{Hioe1984} F. T. Hioe, Phys. Rev. A \textbf{30}, 2100 (1984).

\bibitem{Silver1985} M. S. Silver, R. I. Joseph, and D. I. Hoult, Phys. Rev. A \textbf{31}, 2753 (1985).

\bibitem{Bambini1981} A. Bambini and P. R. Berman, Phys. Rev. A \textbf{23}, 2496 (1981).

\bibitem{Vasilev2004} G. S. Vasilev and N. V. Vitanov, Phys. Rev. A \textbf{70}, 053407 (2004).

\bibitem{Vasilev2005} G. S. Vasilev and N. V. Vitanov, J. Chem. Phys. \textbf{123}, 174106 (2005).

\bibitem{Landau1932} L. D. Landau, Phys. Z. Sowjetunion \textbf{2}, 46 (1932).

\bibitem{Zener1932} C. Zener, Proc. R. Soc. A \textbf{137}, 696 (1932).

\bibitem{Stueckelberg1932} E. C. G. St{\"u}ckelberg, Helv. Phys. Acta \textbf{5}, 369 (1932).

\bibitem{Majorana1932} E. Majorana, Nuovo Cimento \textbf{9}, 43 (1932).

\bibitem{Vitanov1996} N. V. Vitanov and B. M. Garraway, Phys. Rev. A \textbf{53}, 4288 (1996); Erratum Phys. Rev. A \textbf{54}, 5458(E) (1996).


\bibitem{Mokhberi2020}
A. Mokhberi, M. Hennrich, F. Schmidt-Kaler, Adv. At. Mol. Opt. Phys.\textbf{69}, 233 (2020).

\bibitem{Saffman2010} M. Saffman, T. G. Walker, and K. M{\o}lmer, Rev. Mod. Phys. \textbf{82}, 2313 (2010).

\bibitem{Fan2015}
H. Fan, S. Kumar, J. Sedlacek, H. K\"ubler, S. Karimkashi, and J. P. Shaffer, J. Phys. B: At. Mol. Opt. Phys. \textbf{48}, 202001 (2015).

\bibitem{Adams2020} C. S. Adams, J. D. Pritchard, J. P. Shaffer, J. Phys. B: At. Mol. Opt. Phys. \textbf{53}, 012002 (2020).

\bibitem{Carr1954} H. Y. Carr and E. M. Purcell, Phys. Rev. \textbf{94}, 630 (1954).

\bibitem{Meiboom1958} S. Meiboom and D. Gill, Rev. Sci. Instrum. \textbf{29}, 688 (1958).

\bibitem{Maudsley1986} A. A. Maudsley, J. Magn. Reson. \textbf{69}, 488 (1986).

\bibitem{Viola1999} L. Viola, E. Knill, and S. Lloyd, Phys. Rev. Lett. \textbf{82}, 2417 (1999).

\bibitem{Gullion1990} T. Gullion, D. B. Baker, and M. S. Conradi, J. Magn. Reson. \textbf{89}, 479 (1990).

\bibitem{Khodjasteh2005} K. Khodjasteh and D. A. Lidar, Phys. Rev. Lett. \textbf{95}, 180501 (2005).

\bibitem{Piltz2013} C. Piltz, B. Scharfenberger, A. Khromova, A. F. Varon, and C. Wunderlich, Phys. Rev. Lett. \textbf{110}, 200501 (2013).

\bibitem{Casanova2015} J. Casanova, Z.-Y. Wang, J. F. Haase, and M. B. Plenio, Phys. Rev. A \textbf{92}, 042304 (2015).

\bibitem{Souza2011} A. M. Souza, G. A. Alvarez, and D. Suter, Phys. Rev. Lett. \textbf{106}, 240501 (2011).


\bibitem{Genov2017} G. T. Genov, D. Schraft, N. V. Vitanov, and T. Halfmann, Phys. Rev. Lett. \textbf{118}, 133202 (2017).

\bibitem{Vitanov2015} N. V. Vitanov, Phys. Rev. A \textbf{92}, 022314 (2015).

\bibitem{Souza2012} A. Souza, G. A. Alvarez, and D. Suter, Phil. Trans. R. Soc. A \textbf{370}, 4748 (2012).



\end{thebibliography}
\end{document}